\documentclass[12pt]{article}
\usepackage{epsfig,amsfonts,amssymb}
\begin{document}

\newcommand{\be}{\begin{equation}}
\newcommand{\ee}{\end{equation}}

\newcommand{\ba}{\begin{eqnarray}}
\newcommand{\ea}{\end{eqnarray}}
\newcommand{\beas}{\begin{eqnarray*}}
\newcommand{\eeas}{\end{eqnarray*}}

\baselineskip 14 pt

\begin{titlepage}
\begin{flushright}
{\small }
\end{flushright}

\begin{center}

\vspace{5mm}

{\Large \bf Late time evolution of brane gas cosmology and compact
internal dimensions}

\vspace{3mm}

Jin Young Kim \footnote{\tt jykim@kunsan.ac.kr}

\vspace{2mm}

{\small \sl Department of Physics, University of California,
Davis, CA 95616 }

\centerline{and}

{\small \sl Department of Physics, Kunsan National University,
Kunsan, Chonbuk 573-701, Korea}

\date{\today}

\end{center}

\vskip 0.3 cm

\noindent We study the late-time behavior of a universe in the
framework of brane gas cosmology. We investigate the evolution of
a universe with a gas of supergravity particles and a gas of
branes. Considering the case when different dimensions are
anisotropically wrapped by various branes, we have derived
Friedman-like equations governing the dynamics of wrapped and
unwrapped subvolumes. We point out that the compact internal
dimensions are wrapped by three or higher dimensional branes.

\vskip 0.5 cm


\end{titlepage}

\section{Introduction}

Unifying gravity with other forces of nature strongly suggests
that there may be more than three spatial dimensions in the
unifying scale. Some of these are hidden from the low energy
observers. This is related to the cosmological question why we
have only three spatial dimensions. There had been earlier
attempts to answer this question \cite{choddet, qmfl, infl, lz,
bv}. One of the notable is the idea of string cosmology proposed
by Brandenberger and Vafa (BV) \cite{bv}. This is a mechanism to
generate dynamically the spatial dimensionality of spacetime and
to explain the problem of initial singularity. The key ingredient
of this model is based on the symmetry of string theory called
T-duality. With this symmetry the spacetime has a topology of nine
dimensional torus and its dynamics is driven by a gas of
fundamental strings. Cosmology based on this string-modified
Einstein-Friedman equations was studied in many directions
\cite{tv, vene, cleros, msake}.

Developments in string theory during the last decade revealed that
it is not a theory of only strings and has richer structure of
branes \cite{polch}. Recently there have been attempts to
understand the cosmological evolution in this new framework of
string theory with D-branes \cite{magrio, psl} and the mechanism
of BV was considered in \cite{abe}. This model of brane gas
cosmology was studied extensively
\cite{bgextended,alex,egjk1,cbbc}. In the picture of brane gas
cosmology, the universe starts from hot, dense gas of D-branes in
thermal equilibrium. The winding modes of branes obstruct the
growing of the spatial dimensions. Branes with opposite winding
numbers can annihilate if their world volume interacts. Thus
hierarchy of scales can be achieved between the wrapped and
unwrapped dimensions. Of particular interest is whether the
unwrapped configuration of branes can successfully inflate to make
the unwrapped dimensions grow indefinitely while making the
wrapped ones remain small or at least grow much slowly
\cite{egjk1,cbbc}.

In \cite{egjk1}, the authors studied the late-time cosmology in
M-theory with a supergravity particle gas and wrapped 2-brane gas.
In this paper, we study the late-time behavior of brane gas
cosmology in the context of string theory by extending the
formalism of \cite{egjk1} to general $p$-brane gas. We investigate
the behavior of a universe with a gas of supergravity particle and
a gas of wrapped $p$-brane. Considering the case where different
directions are anisotropically wrapped by various branes, we will
argue the possible hierarchical evolution of scales between
wrapped and unwrapped dimensions.

\section{Brane gas dynamics}

In this section, we extend the formalism of brane gas dynamics in
\cite{egjk1} for branes of arbitrary dimensionality and set up
some preliminaries for our calculation. The model we will consider
is type II string theory compactified on $T^9$. Let us consider
the late stage of BV scenario where the radii and curvature scales
are grown larger than the ten-dimensional Planck length. When the
radii are grown enough, we can neglect the brane-antibrane
annihilation. And the excitations on the branes will be
red-shifted away faster than the brane tension. Supergravity is a
good approximation with the growing radii and falling temperature.
Matter fields can be classified into two types. One is the
massless supergravity particles corresponding to bosonic and
fermionic degrees of freedom. We ignore massive modes since these
will decay quickly. The other is wrapped D-branes. These branes do
not interact with each other when the radii are grown large. We
also ignore the fluctuations within the brane. We assume that the
supergravity particles and brane gases are homogeneous for
simplicity of calculation. Then we can take the averages of the
contributions of all kinds of particles and branes to the
energy-momentum tensor.

In the point of string theory, the gravitational interaction is
described by the coupled system of the metric and dilaton. Dilaton
plays an important role in the large-small symmetry of string
theory called T-duality. To be specific, let us start from the
following effective action of type II string theory

 \be
S= \int d^{10}x \sqrt{- g^S} e^{-2 \phi}  \Bigl[R + 4 (\nabla
\phi)^2 \Bigr] + S_m , \label{eaStr}
 \ee
 where $\phi$ is the dilaton field and $S_m$ denote all
 matter actions including the brane action.
 Since we are interested in the late time cosmology,
 we work in the Einstein frame, defined by

 \be
 g_{MN}^{S} = e^{\frac{1}{2} \phi} g_{MN}^{E}.
 \ee
In terms of the Einstein metric, the action can be written as

 \be
S = \int d^{10}x \sqrt{-g^E} \Bigl[ R -\frac{1}{2} (\nabla \phi)^2
\Bigr] + S_m . \label{eaEin}
 \ee
 We drop the superscript $E$ from now on. The equations of motion
 from this action are

 \be  \label{eeqgrav}
 R_{MN} - {1 \over 2} g_{MN} R = {1 \over 2} \nabla_M \phi \nabla_N \phi
       -  {1 \over 4} g_{MN} (\nabla \phi)^2 -
   {1 \over \sqrt{-g} } { {\delta S_m} \over {\delta g^{MN}} },
 \ee
 \be  \label{eeqdil}
  \nabla^2 \phi =
      - {1 \over \sqrt{-g} } { {\delta S_m} \over {\delta \phi} }.
 \ee
Since the string coupling ($g \equiv e^\phi$) is considered to be
small in our assumption of large radii, we do not consider the
running of the dilaton throughout this paper. Then the equations
of motion are described simply by the Einstein equation

 \be
 R_{MN} - {1 \over 2} g_{MN} R =  -
   {1 \over \sqrt{-g} } { {\delta S_m} \over {\delta g^{MN}} } .
 \ee

With the metric ansatz of torus, with $D=9$,

 \be
 ds^2 = - dt^2 + \sum_{i=1}^D \bigl(a_i(t)\bigr)^2
 d\theta_i^2 , \qquad 0 \leq \theta_i \leq 2\pi ,
 \ee
the non-vanishing components of the Einstein tensor are

 \ba
G^t{}_t & = & {1 \over 2} \sum_{k \not= l} {\dot{a}_k \dot{a}_l
\over a_k a_l},  \\
G^i{}_i & = & \sum_{k \not= i} {\ddot{a}_k \over a_k} + {1 \over
2} \sum_{k \not= l} {\dot{a}_k \dot{a}_l \over a_k a_l} - \sum_{k
\not= i} {\dot{a}_k \dot{a}_i \over a_k a_i},
 \ea
where $i$ is not summed in the second equation.

For the matter part, we first consider a gas of massless
supergravity particles, with energy density $\rho_{\rm S}$ and
pressure $p_{\rm S}$.  Since we assume the gas to be homogeneous
and isotropic, we take a perfect fluid form of energy momentum
tensor

 \be
 T^M{}_N = {\rm diag}(-\rho_{\rm S},p_{\rm S},\ldots,p_{\rm S}).
 \ee
The equation of state appropriate for $D$ spatial dimensions fixes
$p_{\rm S} = (1 /D) \rho_{\rm S}$.

The second source of energy momentum comes from a gas of branes,
wrapped on the various cycles of the torus. The matter
contribution of a single $p$-brane to the action, in string frame,
is represented by the Dirac-Born-Infeld (DBI) action

 \be
S_{\rm p} = - T_p \int d^{p+1} \xi e^{-\phi} \sqrt{ - {\rm det} (
{ g}_{\mu\nu} + { B}_{\mu\nu} + 2 \pi \alpha^\prime {F}_{\mu\nu} )
} , \label{pbract}
 \ee
where $T_p$ is the tension of $p$-brane and $g_{\mu\nu}$ is the
induced metric to the brane

 \be
 {g}_{\mu\nu} = g_{MN} \frac{\partial X^M}{ \partial \xi^\mu}
\frac{\partial X^N}{ \partial \xi^\nu}.
 \ee
Here $M,N$ are the indices of $(D+1)$ dimensional bulk spacetime
and $\mu,\nu$ are those of brane. ${ B}_{\mu\nu}$ is the induced
antisymmetric tensor field and ${F}_{\mu\nu}$ is the field
strength tensor of gauge fields $A_\mu$ living on the brane. With
the assumption of grown radii, we neglect ${ B}_{\mu\nu}$ and
${F}_{\mu\nu}$ terms below. Ignoring the running of the dilaton,
we can absorb the effect of constant dilaton into the redefinition
of brane tension in the Einstein frame.

Let us consider a gas of 1-branes. The contribution of these
branes are characterized by wrapping numbers $N_i$ and ${\bar
N}_i$, where we take $N_i$ to represent the number of 1-branes
wrapped on the $i$ cycle, while ${\bar N}_i$ represents the number
of anti-1-branes. By symmetry it is a reasonable assumption that
they are equal. The wrapping numbers are based on the
understanding of thermal fluctuations in the early universe. A
single 1-brane action is described by

 \be
 S_{\rm 1-brane} =  - T_1 \int d^2 \xi \sqrt{- \det g_{\alpha\beta}} ,
 \ee
where $T_1$ is the 1-brane tension. The stress tensor from this
1-brane action is

 \be
 T^{MN} = - T_1 \int d^2\xi \delta^{D+1}(X - X(\xi))
 \sqrt{- \det g} \, g^{\alpha\beta} \partial_\alpha X^M \partial_\beta
 X^N .
 \ee
As an example, for a single 1-brane wrapped on the $1$ cycle and
uniformly smeared over the transverse $T^{D-1}$, the stress energy
tensor is

 \be
 T^M{}_N = - {T_1 \over 2 \pi a_2 \cdots 2 \pi a_D} {\rm
 diag}(1,1,0,0,\ldots,0) .
 \ee
With wrapping numbers $N_i$ and ${\bar N}_i$ for $i$ cycle, the
non-zero components of the 1-brane gas stress tensor are

 \ba
 T^t{}_t & = & - {T_1 \over V} 2\pi \sum_{k} a_k ( N_k + {\bar N}_k ) , \\
 T^i{}_i & = & - {T_1 \over V} 2\pi a_i ( N_i + {\bar N}_i ) .
 \ea

Now we insert these two sources of energy-momentum into the right
hand side of the Einstein equations,

 \be
 G^M{}_N = - 8 \pi G T^M{}_N .
 \ee
The time component equation can be solved for the energy density
of the supergravity gas,

 \be  \label{constr-1br}
 8 \pi G \rho_{\rm S} = {1 \over 2} \sum_{k \not=l} {\dot{a}_k
 \dot{a}_l \over a_k a_l} - {8 \pi G T_1 \over V} (2\pi) \sum_{k}
 a_k (N_k + {\bar N}_k ) .
 \ee
The space component equations, after some algebra, can be reduced
to the following set of second-order differential equations

 \ba
{\ddot{a}_i \over a_i} & = & {8 \pi G T_1 \over V} \left[{D+1
\over D(D-1)} 2\pi \sum_{k} a_k (N_k + {\bar N}_k ) - 2\pi a_i
\left(N_i + {\bar N}_i\right) \right]
\nonumber \\
& & + {1 \over 2 D} \sum_{k \not= l} {\dot{a}_k \dot{a}_l \over
a_k a_l} - \sum_{k \not= i} {\dot{a}_k \dot{a}_i \over a_k a_i} .
\label{eom-1br}
 \ea


For a gas of 2-branes, the wrapping can be characterized by
$N_{ij}$, where we take $N_{i<j}$ to represent the number of
2-branes wrapped on the $(ij)$ cycle, while $N_{i>j}$ represents
the number of anti-2-branes. The stress energy tensor from a
single 2-brane action is

 \be
 T^{MN} = - T_2 \int d^3\xi \delta^{D+1}(X - X(\xi))
 \sqrt{- \det g} \, g^{\alpha\beta} \partial_\alpha X^M \partial_\beta
 X^N ,
 \ee
where $T_2$ is the 2-brane tension. Then, for a single 2-brane
wrapped on the $(12)$ cycle and uniformly smeared over the
transverse $T^{D-2}$, the stress energy tensor is

 \be
 T^M{}_N = - {T_2 \over 2 \pi a_3 \cdots 2 \pi a_D} {\rm
 diag}(1,1,1,0,\ldots,0) .
 \ee
With the wrapping numbers, the non-zero components of stress
energy tensor from the 2-brane gas are

 \ba
T^t{}_t & = & - {T_2 \over V} (2\pi)^2 \sum_{k \not= l} a_k  a_l N_{kl} , \\
T^i{}_i & = & - {T_2 \over V} (2\pi)^2 \sum_{k \not= i} a_k  a_i
\left(N_{ki} + N_{ik}\right) .
 \ea
Repeating the same procedure as in 1-brane case, we have

 \be
\label{constr-2br} 8 \pi G \rho_{\rm S} = {1 \over 2} \sum_{k
\not=l} {\dot{a}_k \dot{a}_l \over a_k a_l} - {8 \pi G T_2 \over
V} (2\pi)^2 \sum_{k \not= l} a_k  a_l N_{kl} ,
 \ee

 \ba
 {\ddot{a}_i \over a_i} & = & {8 \pi G T_2 \over V}
\left[{2D+1 \over D(D-1)} \, (2\pi)^2 \sum_{k \not= l}  a_k  a_l
N_{kl} - (2\pi)^2 \sum_{k \not= i}  a_k  a_i \left(N_{ki} +
N_{ik}\right) \right]
\nonumber \\
& & + {1 \over 2 D} \sum_{k \not= l} {\dot{a}_k \dot{a}_l \over
a_k a_l} - \sum_{k \not= i} {\dot{a}_k \dot{a}_i \over a_k a_i} .
\label{eom-2br}
 \ea


For a gas of 3-branes, we characterize the wrapping numbers of
$(ijk)$ cycle as $N_{ijk}$, where we take $N_{ijk}$ with
$\epsilon_{ijk} = 1$ to represent the number of 3-branes wrapped
on the $(ijk)$ cycle, while $N_{ijk}$ with $\epsilon_{ijk} = - 1$
represents the number of anti-3-branes. The stress energy tensor
for a single 3-brane action is

 \be
 T^{MN} = - T_3 \int d^4\xi \delta^{D+1}(X - X(\xi))
 \sqrt{- \det g} \, g^{\alpha\beta} \partial_\alpha X^M \partial_\beta
 X^N ,
 \ee
 where $T_3$ is the 3-brane tension.
For a single 3-brane wrapped on the $(123)$ cycle and uniformly
smeared over the transverse $T^{D-3}$, the stress energy tensor is

 \be
 T^M{}_N = - {T_3 \over 2 \pi a_4 \cdots 2 \pi a_D} {\rm
 diag}(1,1,1,1,0,\ldots,0) .
 \ee
The non-zero components of the 3-brane gas stress energy tensor
are

 \ba
T^t{}_t & = & - {T_2 \over V} (2\pi)^3 \sum_{k \not= l \not=m}
 {1 \over 3} a_k a_l a_m N_{klm} , \\
T^i{}_i & = & - {T_2 \over V} (2\pi)^3 \sum_{k \not= l \not= i} {1
\over 3} a_k a_l a_i \left(N_{ikl} + N_{lik}+ N_{kli} \right)
 ,
 \ea
where $\sum_{k \not= l \not=m}$ means that the sum is taken when
all three indices are different. The constraint equation and the
second-order differential equations for the radii are

 \be
\label{constr-3br} 8 \pi G \rho_{\rm S} = {1 \over 2} \sum_{k
\not=l} {\dot{a}_k \dot{a}_l \over a_k a_l} - {8 \pi G T_3 \over
V} (2\pi)^3 \sum_{k \not= l \not= m} {1 \over 3} a_k a_l a_m
N_{klm} ,
 \ee

 \ba
 {\ddot{a}_i \over a_i} & = & {8 \pi G T_3 \over V}
 \Big[ {3D+1 \over D(D-1)} \, (2\pi)^3 \sum_{k \not= l \not= m}
 {1 \over 3} a_k a_l a_m N_{klm}   \nonumber  \\
  & &~~~~~~~~~~ - (2\pi)^3
 \sum_{k \not= l \not= i} {1 \over 3} a_k a_l a_i
\left(N_{ikl} + N_{lik} + N_{kli} \right) \Big ]
\nonumber \\
& & + {1 \over 2 D} \sum_{k \not= l} {\dot{a}_k \dot{a}_l \over
a_k a_l} - \sum_{k \not= i} {\dot{a}_k \dot{a}_i \over a_k a_i} .
\label{eom-3br}
 \ea

\section{Late time behavior}
To study the solutions of the field equations (\ref{eom-1br}),
(\ref{eom-2br}) and (\ref{eom-3br}), we introduce new variables
 $\lambda_i(t) \equiv \ln{(2 \pi a_i(t))}$ as in \cite{tv}.
 The field equations when the matter fields are coming from
 supergravity particles and wrapped 1-brane gas become,
 from (\ref{eom-1br}),

 \be
 \label{eoml-1br} \ddot{\lambda_i} + {\dot{\rm V} \over {\rm V}}
\dot{\lambda}_i = 8 \pi G \left({1 \over D} \rho_{\rm S} + {2
\over D - 1} \rho_{\rm B}\right) - {8 \pi G T_1 \over V}
e^{\lambda_i} ( N_i + {\bar N}_i ) ,
 \ee
where the volume of the spatial torus $V$ can be expressed as
 $ V = e^{\sum_i \lambda_i} $
 and $\rho_{\rm B}$ is the energy density from 1-brane tension

 \be
 \rho_{\rm B} = {T_1 \over V} \sum_{k}
 e^{\lambda_k} ( N_i + {\bar N}_k ) .
 \ee
The energy density from supergravity particles is fixed by the
constraint equation (\ref{constr-1br})

 \be
 \label{constrl-1br}
 8 \pi G \rho_{\rm S} = {1 \over 2} \sum_{i \not= j}
 \dot{\lambda_i} \dot{\lambda_j} - 8 \pi G \rho_{\rm B}.
 \ee

This system of equations can be regarded as a non-relativistic
particle moving in $D$ dimensions. The particle has a coefficient
of friction, given by $\dot{V} / V$, due to the expansion of the
universe. One can consider the right hand side of (\ref{eoml-1br})
gives two position-dependent forces acting on the particle
\cite{egjk1}. The first term from the supergravity particles,

 \be
 \label{F1-ibr}
 F_i^{(1)} = 8 \pi G \left({1 \over D} \rho_{\rm S}
 + {2 \over D - 1} \rho_{\rm B}\right) ,
 \ee
 is positive definite.
This force is common to every value of $i$ and drives a uniform
expansion of the universe. The second term from the brane gas
contribution,

 \be
 \label{F2-1br}
 F_i^{(2)} = - {8 \pi G T_1 \over V}
 e^{\lambda_i} (N_i + {\bar N}_i) ,
 \ee
can be either zero or negative depending on the wrapping number.
This term can suppress the growth of dimensions when wrapping
numbers are nonzero. Thus anisotropic wrapping numbers can lead to
an anisotropic expansion of the dimensions.

The constraint equation (\ref{constrl-1br}) can be rewritten as

 \be
 \left({\dot{V} \over V}\right)^2 = \left(\sum_i
 \dot{\lambda}_i\right)^2 = \sum_i (\dot{\lambda}_i)^2 + 16 \pi G
 \left(\rho_{\rm S} + \rho_{\rm B}\right) .
 \ee
The right hand side is positive definite for all nontrivial cases
and we choose the direction of time to make $\dot{V} > 0$.

We are interested in cases where the wrapping number $N_i$ is
anisotropic. We can classify the spatial dimensions into two
kinds. We refer to a direction $i$ as {\em unwrapped\/} if $N_i=
{\bar N}_i =0$ and as {\em wrapped\/} if $N_i= {\bar N}_i \not=
0$. Now consider the case when $m$ dimensions are unwrapped and
$D-m$ dimensions are wrapped. This configuration can be achieved
by thermal fluctuations in the early universe through the
mechanism of Brandenberger and Vafa \cite{bv}. Denote the two
subvolumes as in \cite{egjk1}

 \ba
 \mu & = & \sum_{i = 1}^m \lambda_i = \ln \,\, (\hbox{\rm
 volume of unwrapped torus}), \\
 \Lambda & = & \sum_{i = m+1}^D \lambda_i = \ln \,\, (\hbox{\rm
 volume of wrapped torus}) .
 \ea
Summing over the appropriate values of $i$, and using the
definition of $\rho_{\rm B}$, we find the following differential
equations for $\mu$ and $\Lambda$

 \ba
 \label{mu-1br}
 \ddot{\mu} + (\dot{\mu} + \dot{\Lambda}) \dot{\mu}
 & = & 8 \pi G \left({m \over D} \rho_{\rm S} + {2 m \over D-1}
 \rho_{\rm B}\right) , \\
 \label{Lambda-1br}
 \ddot{\Lambda} + (\dot{\mu} + \dot{\Lambda})
 \dot{\Lambda} & = & 8 \pi G \left({D-m \over D} \rho_{\rm S} + {D
 - 2 m + 1 \over D-1} \rho_{\rm B}\right) .
 \ea


Repeating the same procedure when the matter fields are coming
from supergravity particles and wrapped 2-brane gas, we have,
 from (\ref{eom-2br}),

 \be
 \label{eoml-2br}
 \ddot{\lambda_i} + {\dot{\rm V} \over {\rm V}}
\dot{\lambda}_i = 8 \pi G \left({1 \over D} \rho_{\rm S} + {3
\over D - 1} \rho_{\rm B}\right) - {8 \pi G T_2 \over V}
e^{\lambda_i} \sum_{j \not= i} (N_{ij} + N_{ji}) e^{\lambda_j} .
 \ee
The constraint equation has the same form as (\ref{constrl-1br})
with the energy density $\rho_{\rm B}$ replaced  by

 \be
 \rho_{\rm B} = {T_2 \over V} \sum_{i \not= j} e^{\lambda_i}
 e^{\lambda_j} N_{ij} .
 \ee
In this case we can classify the spatial dimensions into three
classes \cite{egjk1}. A direction $i$ is referred to as
 {\em unwrapped\/} if $N_{ij} = N_{ji} = 0$ for all $j$.
Directions $i$ for which $N_{ij}$ and $N_{ji}$ are nonzero except
for those $j$ corresponding to an unwrapped direction are referred
to as {\em fully wrapped\/}. Directions $i$ where some of the
$N_{ij}$ or $N_{ji}$ are zero for values of $j$ which are not
unwrapped are referred to as {\em partially wrapped\/}. The
equations governing the motion of $\mu$ and $\Lambda$ are

 \ba
 \label{mu-2br}
 \ddot{\mu} + (\dot{\mu} + \dot{\Lambda}) \dot{\mu} & =
& 8 \pi G \left({m \over D} \rho_{\rm S} + {3 m \over D-1}
\rho_{\rm B}\right) , \\
\label{Lambda-2br} \ddot{\Lambda} + (\dot{\mu} + \dot{\Lambda})
\dot{\Lambda} & = & 8 \pi G \left({D-m \over D} \rho_{\rm S} + {D
- 3 m + 2 \over D-1} \rho_{\rm B}\right) .
 \ea

Similarly when the matter fields are coming from
 supergravity particles and wrapped 3-brane gas, we have,
 from (\ref{eom-3br}),

 \ba
 \label{eoml-3br}
 \ddot{\lambda_i} + {\dot{\rm V} \over {\rm V}}
 \dot{\lambda}_i &=& 8 \pi G \left({1 \over D} \rho_{\rm S} + {4
 \over D - 1} \rho_{\rm B}\right)  \nonumber  \\
 &-& {8 \pi G T_3 \over V}
 e^{\lambda_i} \sum_{j \not= k \not= i} {1 \over 3} e^{\lambda_j}
 e^{\lambda_k} (N_{ijk} + N_{kij} + N_{jki} ).
 \ea
The constraint equation has the same form as (\ref{constrl-1br})
with the brane energy density $\rho_{\rm B}$ replaced by

 \be
 \rho_{\rm B} = {T_3 \over V} \sum_{i \not= j \not= k} {1 \over 3}
 e^{\lambda_i} e^{\lambda_j} e^{\lambda_k}  N_{ijk} .
 \ee
Also we can classify the spatial dimensions into three classes as
in 2-brane case; {\em unwrapped\/}, {\em partially wrapped\/} and
{\em fully wrapped\/}. The equations for $\mu$ and $\Lambda$ are

 \ba
 \label{mu-3br}
 \ddot{\mu} + (\dot{\mu} + \dot{\Lambda}) \dot{\mu} & =
& 8 \pi G \left({m \over D} \rho_{\rm S} + {4 m \over D-1}
\rho_{\rm B}\right) , \\
\label{Lambda-3br} \ddot{\Lambda} + (\dot{\mu} + \dot{\Lambda})
\dot{\Lambda} & = & 8 \pi G \left({D-m \over D} \rho_{\rm S} + {D
- 4 m + 3 \over D-1} \rho_{\rm B}\right) .
 \ea

 Generalizing the above result to the case when matter fields are
 coming from supergravity particles and wrapped $p$-brane gas, one
 can find

 \ba
 \label{mu-pbr}
 \ddot{\mu} + (\dot{\mu} + \dot{\Lambda}) \dot{\mu} & =
& 8 \pi G \left({m \over D} \rho_{\rm S} + { {(p+1)m} \over D-1}
\rho_{\rm B}\right) , \\
\label{Lambda-pbr}
 \ddot{\Lambda} + (\dot{\mu} + \dot{\Lambda})
\dot{\Lambda} & = & 8 \pi G \left({D-m \over D} \rho_{\rm S} + {D
- (p+1) m + p \over D-1} \rho_{\rm B}\right) .
 \ea
The evolution of unwrapped subvolume $\mu$ and wrapped subvolume
$\Lambda$ depends on the number of unwrapped dimensionality $m$.
For small $m$, both terms on the right hand side of
$\ddot{\Lambda}$ equation (\ref{Lambda-pbr}) are positive. If
$\dot{\Lambda}$ is zero, then the second derivative must be
positive, leading to a local minimum. Conversely, for a larger
value of $m$, the right hand side of this equation has both a
positive and a negative term. In this case both local maxima and
minima are possible. However, for unwrapped subvolume
$\ddot{\mu}$, there are only positive terms on the right hand
side. So if these directions are initially expanding they will
expand forever.

When the second term of right hand side in $\ddot{\Lambda}$
equation vanishes the brane tension does not contribute to the
growth of the internal dimensions. This gives a criterion for the
critical spatial dimensionality of the unwrapped subspace

 \ba
 D- 2m_c + 1 &=& 0 ,~~~~~~ {\rm for} ~~{\rm 1-brane}, \nonumber \\
 D- 3m_c + 2 &=& 0 ,~~~~~~ {\rm for} ~~{\rm 2-brane}, \\
 D- 4m_c + 3 &=& 0 ,~~~~~~ {\rm for} ~~{\rm 3-brane}.  \nonumber
 \ea
For $D=9$, this occurs at $m_c =5, 11/3$ and $3$ for 1-, 2- and
3-brane gases respectively. From this result we can conclude that
wrapping the internal dimensions with 1- and 2-branes cannot yield
the observed three large spatial dimensions. For 3-brane there is
no contribution from $\rho_{\rm B}$ but $\Lambda$ will grow by
$\rho_{\rm S}$. However the growing rate is smaller than that of
$\mu$. When $\mu$ is wrapped by $p$-dimensional brane gas, the
criterion is

 \be
 D - (p+1)m_c + p = 0 ,~~~~~~ {\rm for} ~~ p{\rm -brane}.
 \ee
As an example, for $p=4$, this gives $m_c=13/5$. This means that
when $m > 13/5 $ there can be negative contribution from the
$\rho_{\rm B}$ terms. Our analysis implies that the compact extra
dimensions should be wrapped with three or higher dimensional
branes for the observed three large spatial dimensions. If they
are wrapped by just one or two dimensional branes, they cannot
remain compact.

\section{Conclusion and Discussion}

We have studied the late-time behavior of the universe in the
framework of brane gas cosmology. For the cases when different
dimensions are anisotropically wrapped by various branes, we have
derived Friedman-like equations governing the dynamics of wrapped
and unwrapped subvolumes. We pointed out that one cannot keep the
extra dimensions compact with one or two dimensional brane gases.

Though the wrapped directions grow slowly compared with the
unwrapped dimensions, it is not guaranteed that the compact
dimensions can be stabilized. If four or higher dimensional branes
are frozen at the end of thermal stage, these branes give negative
force to $\Lambda$ while giving positive force to $\mu$. If we
assume that every term in $\rho_{\rm B}$ takes a comparable
contribution to the energy density, we can replace the effect of
all wrapped branes with a ($D-m$)-brane gas. In this point of
view, one can study the evolution of a universe starting from the
metric

 \be
 g_{MN} = {\rm diag} ( -1, a^2 \delta_{ij} , b^2 \delta_{mn} ),
 \ee
where $a^2$ is the scale factor of the three dimensional space and
$b$ is the scale factor of the internal $D-3$ dimensional
subspace. The stabilization of the extra dimensions with this
asymmetric setting was attempted in the brane world scenario
\cite{adkmr}. Recently the stabilization of extra dimensions was
studied numerically. In string gas cosmology, cosmological
solution for the late time evolution showed that large dimensions
continue to expand and the small ones are kept undetectably small
\cite{borunda}. In \cite{brwa}, it is shown that six compact
dimensions become stabilized at the self-dual radius while three
dimensions grow large. More recently it has been shown that, by
reducing the effect of string gas to the four dimensional Einstein
gravity, string modes cannot stabilize the internal dimensions
wrapped by winding strings, except in the special case of one
extra dimension \cite{batwat}. However, the evolution of extra
dimensions is slow enough compared to the unwrapped large
dimensions.

One way to achieve the stability of the internal dimensions is to
include the effect of fluxes. Alexander \cite{alex} pointed out
the possibility of stabilizing the extra dimensions with G-flux in
the context of brane gases in eleven dimensional supergravity
derived from M-theory. We expect similar argument can be applied
to brane gases in ten dimensional supergravity from type II string
theory. For example, if we consider the effect of NS-NS sector in
the bulk action, the cosmology is described by

 \be
S= \int d^{10}x \sqrt{- g^S} e^{-2 \phi}  \Bigl[R + 4 (\nabla
\phi)^2 - \frac{1}{12} H_{LMN}^2  \Bigr] + S_m , \label{actwithh}
 \ee
which involves the dynamics of two-form field $B_{MN}$ through
$H_{LMN} = 3 \nabla_{[L} B_{MN ]}$. Then the stability of the
internal dimensions can be studied by the coupled three equations
of motion involving $g_{MN}$, $\phi$ and $H_{LMN}$ instead of two
(Eqs. \ref{eeqgrav} and \ref{eeqdil}). We expect an effective
potential with a confining form so that the internal dimensional
will oscillate and remain small.

 Classical supergravity holds when all radii are
larger than the ten dimensional Planck length. So the equations we
used are valid when the radii are either constant or growing with
time. In this case we can safely neglect the massless excitations
on the branes which will redshift quickly. We can also neglect the
brane antibrane annihilations because branes will be frozen as the
transverse dimensions expand. However, in string theory,
gravitational interaction is described not by metric alone but by
the coupled system of metric and dilaton. The assumption that
dilaton is constant may be inconsistent with the cosmological
equations describing the evolution. Also the production of stringy
objects in the general time dependent background is complicated
since the back reaction of such strings or branes is likely to be
important. Further studies on the late-time behavior of brane gas
cosmology with interacting branes together with dilaton field are
needed.

\vspace{.5cm} {\bf Acknowledgements}

We are grateful to the Department of Physics at U.C. Davis for
hospitality during the completion of this work. We would like to
thank N. Kaloper for suggestion and helpful discussion. We also
thank R. Brandenberger for pointing out the relevance of
\cite{batwat} where some of the main conclusions overlap with
ours. This work was supported by Korea Research Foundation Grant
(KRF-2001-015-DP0082) and Kunsan National University's Long-term
Overseas Research Program for Faculty Member in the year 2003.


\begin{thebibliography}{99}

\bibitem{choddet}
A. Chodos and S. Detweiler, Phys. Rev. {\bf D21}, 2167 (1980).

\bibitem{qmfl}
P. G. O. Freund, Nucl. Phys. {\bf B209}, 146 (1982); T. Appelquist
and A. Chodos, Phys. Rev. {\bf D28}, 772 (1983); T. Appelquist, A.
Chodos and E. Myers, Phys. Lett. {\bf B127}, 51 (1983).

\bibitem{infl}
Q. Shafi and C. Wetterich, Phys. Lett. {\bf B129}, 387 (1983);
Phys. Lett. {\bf B152}, 51 (1985); Y. Tosa, Phys. Rev. {\bf D30},
339 (1984); Phys. Rev. {\bf D30}, 2054 (1984); Y. Okada, Phys.
Lett. {\bf B150}, 103 (1985); Y. Okada and M. Yoshimura, Phys.
Rev. {\bf D33}, 2164 (1986).

\bibitem{lz}
A. D. Linde and M. I. Zelnikov, Phys. Lett. {\bf B215}, 59 (1988).

\bibitem{bv}
R. Brandenberger and C. Vafa, Nucl. Phys. {\bf B316}, 391 (1989).

\bibitem{tv}
A. A. Tseytlin and C. Vafa, Nucl. Phys. {\bf B372}, 443 (1992); A.
A. Tseytlin, Class. Quant. Grav. {\bf 9}, 979 (1992).

\bibitem{vene}
N. Sanchez and G. Veneziano, Nucl. Phys. {\bf B333}, 253 (1990);
G. Veneziano, Phys. Lett. {\bf B265}, 287 (1991); M. Gasperini, N.
Sanchez and G. Veneziano, Nucl. Phys. {\bf B364}, 365 (1991).

\bibitem{cleros} G. B. Cleaver and P. J. Rosenthal,
Nucl. Phys. {\bf B457}, 621 (1995).

\bibitem{msake} M. Sakellariadou, Nucl. Phys. {\bf B468}, 319 (1996).

\bibitem{polch}
J. polchinski, Phys. Rev. Lett. {\bf 75}, 4724 (1995);
E. Witten, Nucl. Phys. {\bf B443}, 85 (1995);
J. polchinski, {\it String theory} (Cambridge University Press,
Cambridge, England, 1998).

\bibitem{magrio}
M. Maggiore and A. Riotto, Nucl.Phys. {\bf B548}, 427 (1999).

\bibitem{psl}
C. Park S.-J. Sin and S. Lee, Phys. Rev. {\bf D61}, 083514 (2000).

\bibitem{abe}
S. Alexander, R. Brandenberger and D. Easson, Phys. Rev. {\bf D62}
 013509 (2000).

\bibitem{bgextended}
R. Brandenberger, D. A. Easson and D. Kimberly, Nucl. Phys. {\bf
B623}, 421 (2002);
D. A. Easson, Int. J. Mod. Phys. {\bf A18}, 4295 (2003);
M. F. Parry and D. A. Steer, JHEP {\bf 0202}, 032 (2002);
R. Easther, B. R. Greene and M. G. Jackson, Phys. Rev. {\bf D66},
023502 (2002);
S. Watson and R. H. Brandenberger, Phys. Rev. {\bf D67}, 043510
(2003);
T. Boehm and R. Brandenberger, JCAP {\bf 0306}, 008 (2003);
A. Kaya and T. Rador, Phys. Lett. {\bf B565}, 19 (2003);
A. Kaya, Class. Quant. Grav. {\bf 20}, 4533 (2003);
R. Easther, B. R. Greene, M. G. Jackson, and D. Kabat, JCAP
 {\bf 0401}, 006 (2004).

\bibitem{alex}
S. H. S. Alexander, JHEP {\bf 0310}, 013 (2003).

\bibitem{egjk1}
R. Easther, B. R. Greene, M. G. Jackson, and D. Kabat, Phys. Rev.
{\bf D67}, 123501 (2003).

\bibitem{cbbc}
A. Campos, Phys. Rev. {\bf D68}, 104017 (2003);
R. Brandenberger, D. A. Easson and A. Mazumdar, Phys. Rev.
 {\bf D69}, 083502 (2004);
T. Biswas, JHEP {\bf 0402}, 039 (2004);
A. Campos, Phys. Lett. {\bf B586}, 133 (2004).

\bibitem{adkmr}
N. Arkani-Hamed S. Dimopoulos, N. Kaloper, and J. March-Russell,
Nucl. Phys. {\bf B567}, 189 (2000);
A. Riotto, Phys. Rev. {\bf D61}, 123506 (2000).

\bibitem{borunda}
B. A. Bassett, M. Borunda, M. Serone and S. Tsujikawa, Phys. Rev.
 {\bf D67} 123506 (2003).

\bibitem{brwa}
S. Watson and R. H. Brandenberger, JCAP {\bf 0311}, 008 (2003).

\bibitem{batwat}
T. Battefeld and S. Watson, JCAP {\bf 0406}, 001 (2004).


\end{thebibliography}
\end{document}